\documentclass{jjap3}

\title{Theoretical limit of power conversion efficiency for organic and hybrid halide perovskite photovoltaics}
\author{Kazuhiko Seki\thanks{E-mail: k-seki@aist.go.jp},  Akihiro Furube$^{1}$, and Yuji Yoshida$^{2}$}
\inst{NMRI,  RIIF$^{1}$,  RCPV$^{2}$, National Institute of Advanced Industrial Science and Technology (AIST),  
Tsukuba, Ibaraki 305-8565  Japan
}
\abst{We calculated the maximum power conversion efficiency as a function of the optical band gap for organic photovoltaic (PV) cells by assuming that charge separation is accompanied by the energy loss required to dissociate strongly bound charge pairs. The dissociation energy can be estimated from the relationship between the open circuit voltage ($V_{\rm OC} $) and the optical band gap ($E_{\rm g}$). By analyzing the published data on $V_{\rm OC}$  and $E_{\rm g}$, the dissociation energy can be estimated. The result could be used as a guide for selecting donor and acceptor materials.
We also studied the theoretical limit of power conversion efficiency of hybrid halide perovskite 
by taking into account the energy loss involved in the carrier transfer from the perovskite phase to 
the metal oxide charge transport layer.
}

\begin{document}
\maketitle

\section{Introduction}
Organic photovoltaic (OPV) cells have many advantages, such as being thin, soft, and light, over conventional inorganic photovoltaic cells owing to characteristics of organic materials. Although the photoelectric power conversion efficiency of organic photovoltaics is lower than that of inorganic photovoltaic cells, it has improved rapidly and now exceeds 10 \%. 
The rapid increase in the power conversion efficiency aroused fundamental interest in the theoretical limit of the conversion efficiency for organic solar cells. 

Since the pioneering work of Shockley and Queisser in 1961 \cite{Shockley}, 
the theoretical limit of the power  conversion efficiency has been known for inorganic solar cells and is approximately 30\%. 
The Shockley--Queisser (SQ)  limit was calculated for PN-junction solar cells and 
is not applicable for excitonic solar cells. 
The exciton-binding energy can be at least as large as $0.3-0.5$ eV and cannot be ignored in organic photovoltaic cells. 
Organic photovoltaic cells are composed of donor and acceptor materials, and charge separation against the large exciton-biding energy occurs at the donor-acceptor interface. 

The application of the SQ limit  has been extended to organic photovoltaics.  \cite{Kirchartz,Giebink,Gruber,Koster,Janssen,Seki,Miyadera,Scharber} 
Some other approaches have also been developed to study the efficiency, in particular, on the basis of a thermodynamic detailed balance. \cite{Einax11,Einax13,Einax,Nelson04,Rutten}
As a practical approach, 
limits for solar cells were assessed using criteria based on the short circuit currents, open circuit voltage and other quantities. 
\cite{Nayak}

When the SQ limit  has been applied to organic photovoltaics,  
the difference between the optical energy gap and the electronic energy gap has been taken into account. \cite{Kirchartz,Giebink,Gruber,Koster,Janssen,Seki,Miyadera,Scharber} 
The energy difference between these two gaps results in voltage loss by energy dissipation. 
Recently, we have calculated the limit of power conversion efficiency by assuming 
irreversible exciton dissociation \cite{Seki}, 
while other studies assumed reversible exciton dissociation. \cite{Kirchartz,Giebink,Gruber,Koster}
We took into account the excess energy required for irreversible charge separation at donor/acceptor (D/A) interfaces. \cite{Seki} 
In this work, we briefly introduce our approach and 
show its consequences with respect to the 
open circuit voltage and the short circuit currents. 

Recently, the power conversion efficiency of hybrid halide perovskite solar cells has been rapidly increased. \cite{Kojima,Im,Kim,Lee,Burschka,Liu,Heo}
The metal oxide TiO$_2$ is used as a charge
transport layer in 
the hybrid halide perovskite photovoltaic cells.  
Although hybrid halide perovskite solar cells are not classified as excitonic solar cells, 
the carrier transfer from the perovskite phase to the metal oxide TiO$_2$ involves intrinsic energy losses.  
We study the power conversion efficiency of hybrid halide perovskite photovoltaics by taking into account 
the energy loss involved in the carrier transfer from the perovskite phase to the metal oxide charge transport layer.

\section{Theoretical limit of organic photovoltaic cells}
\begin{figure}
\centerline{
\includegraphics[width=0.6\columnwidth]{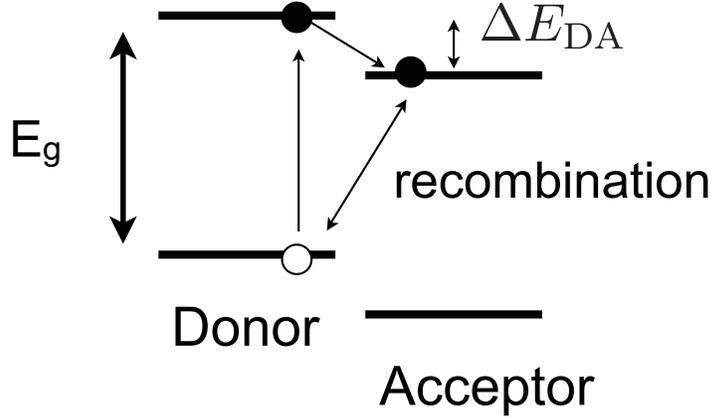}
}
\caption{Schematic illustration of  charge dissociation processes in organic photovoltaic cells. The dissociation energy is denoted by $\Delta E_{\rm DA}$. The optical band gap is denoted by $E_{\rm g}$. 
The figure is reproduced from Ref. \citenum{Seki} with permission. 
}
\label{fig:1}
\end{figure}

In this section, we summarize  the results of our previous work.  \cite{Seki}. 
Organic PV cells are classified as excitonic solar cells, where strongly bound pairs of charge carriers are generated by 
photo-excitations. In our approach, dissociation of excitons into the charge-separated states was considered to be accompanied by a nonradiative dissociation energy. The Coulombic interactions between oppositely charged carriers are strong owing to the low values of dielectric constants of organic materials. As shown schematically in Fig. \ref{fig:1}, charge 
separation takes place at the D/A interface, resulting in the loss of the dissociation energy denoted by $\Delta E_{\rm DA}$.  
          
\begin{figure}
\centerline{
\includegraphics[width=0.6\columnwidth]{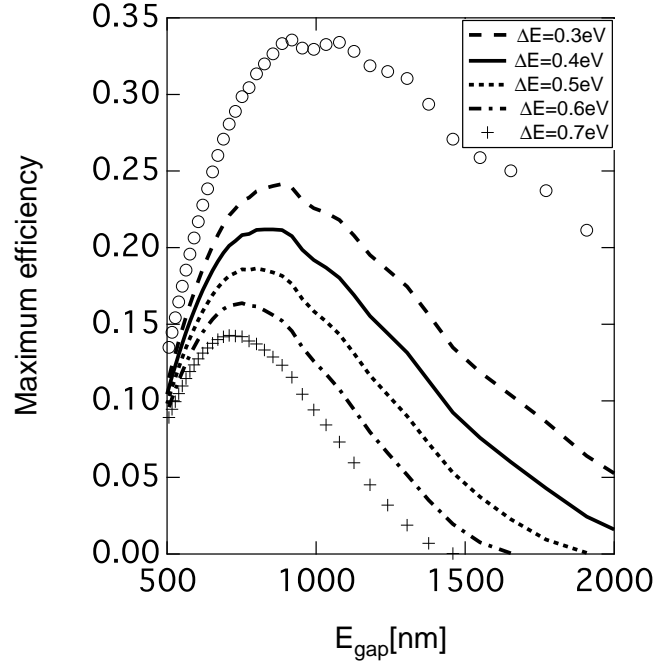}
}
\caption{Maximum power conversion efficiency as a function of the optical band gap (nm). The SQ limit is shown by circles. The other lines indicate the results including the dissociation energy of 
$\Delta E_{\rm DA} =0.3, 0.4, 0.6,\mbox{and } 0.7$ eV from top to bottom.
}
\label{fig:2}
\end{figure}

  The power conversion efficiency is given by the ratio of the maximum electric power to the radiative power irradiated at the solar cell. The input radiative power can be calculated using the AM 1.5 spectrum that we denote by $j_{\rm in}(E)$. 
  For simplicity, we assume that all photons of energy higher than band gap energy are absorbed by the
cell and converted to carriers,  
$J_{\rm in} (E_{\rm g}) = \int_{E_{\rm g}}^\infty d E j_{\rm in} (E)$.   
As in the SQ theory of inorganic PV cells, we assume inevitable loss of carriers by radiative recombination. The loss of carriers by recombination per unit area per unit time can be expressed by \cite{Seki}
\begin{align}
J_{\rm R}  (E_{\rm g} -\Delta E_{\rm DA},V)=\exp \left(\frac{eV}{k_{\rm B}  T}\right) 
\int_{E_{\rm g}-\Delta E_{\rm DA}}^\infty dE  \frac{2\pi E^2}{c^2 h^3} \exp 
\left(-\frac{E}{k_{\rm B}  T} \right),
\label{eq:1}
\end{align}
where $h$, $k_{\rm B}$, and $c$ denote 
the Planck constant, the Boltzmann constant, and the speed of light. 
The maximum power conversion efficiency can be obtained from \cite{Seki}  
\begin{align}
Q(E_{\rm g}, \Delta E_{\rm DA})=\frac{\mbox{Max} \left\{ eV\left[ J_{\rm in}  
(E_{\rm g}  )-J_{\rm R}  (E_{\rm g}- \Delta E_{\rm DA} ,V)\right]\right\}_V}{J_{\rm in} (0) } . 
\label{eq:2}
\end{align}
When 
$\Delta E_{\rm DA} =0$, the above equation reduces to the SQ limit. The maximum power conversion efficiency is shown as a function of the band gap (nm) in Fig. \ref{fig:2}. 
In Fig. \ref{fig:2}, $E_{\rm g}$(eV) is expressed by wave length using  
$E_{\rm gap} (\mbox{nm})=1240 (\mbox{eV$\cdot$ nm})/E_{\rm g} (\mbox{eV})$. 
By increasing $\Delta E_{\rm DA}$, the maximum power conversion efficiency is decreased and the band gap at the peak is shifted 
toward shorter wavelengths. The blue shift of the peak results from the excess energy required to dissociate excitons. 

\section{Dissociation energy}
Plausible values of $\Delta E_{\rm DA}$ can be estimated from the relationship between the open circuit voltage 
and the optical band gap. 
$eV_{\rm OC}$ can be regarded as the electronic band gap. 
By setting $J_{\rm in}  (E_{\rm g} )-J_{\rm R}  (E_{\rm g} -\Delta E_{\rm DA} ,V_{\rm OC})=0$ 
and using the relation $J_{\rm R}  (E ,V)=\exp\left[eV/\left(k_{\rm B}  T \right)\right]J_{\rm R}  (E ,0)$ derived from Eq. (\ref{eq:1}), 
the open circuit voltage is obtained as  
\begin{align}
eV_{\rm OC} = k_{\rm B} T \ln \left[ 
\frac{J_{\rm in} (E_{\rm g}  )}{J_{\rm R}  (E_{\rm g} -\Delta E_{\rm DA} ,0)}
\right]. 
\label{eq:3}
\end{align}
The recombination current density $J_{\rm R}  (E_{\rm g} -\Delta E_{\rm DA} ,0)$ can be expressed by 
the carrier densities at the interface and the recombination life time. 
Essentially the same equation as Eq. (\ref{eq:3}) can be derived in this case, and $V_{\rm OC}$ 
can be expressed using the recombination lifetime. \cite{Credgington}
The results indicate that the recombination lifetime is related directly to $V_{\rm OC}$. \cite{Credgington}

\begin{figure}
\centerline{
\includegraphics[width=0.6\columnwidth]{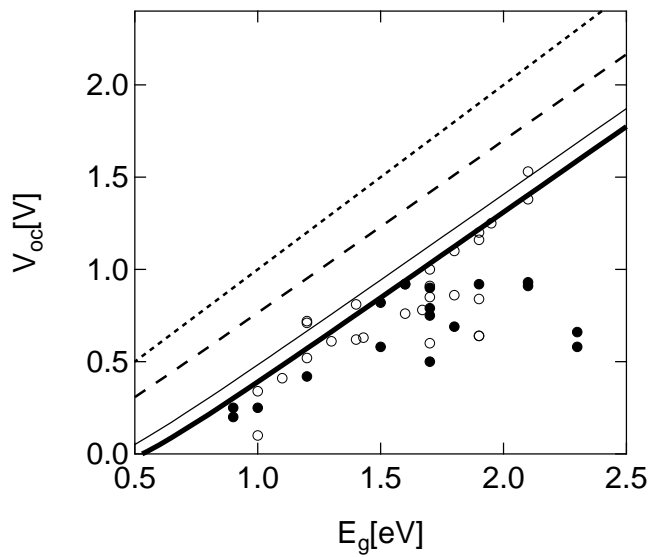}
}
\caption{
Open circuit voltage $V_{\rm OC}$ as a function of the optical band gap (eV) denotes by $E_{\rm g}$. 
The dotted line indicates $e V_{\rm OC}=E_{\rm g}$. The dashed line is obtained from the SQ limit. 
The thin solid line is obtained for $0.3$ eV, and the thick solid line is obtained for 
$\Delta E_{\rm DA}=0.4$ eV. The open circles represent the experimental results obtained using small molecules taken from 
Ref. \citenum{Lunt}. 
The closed circles represent the experimental results obtained using polymers taken from Refs. 
\citenum{Lunt} and \citenum{Veldman}.
}
\label{fig:3}
\end{figure}

In Fig. \ref{fig:3}, we show $V_{\rm OC}$  as a function of  
$\Delta E_{\rm DA}$. 
The result for the SQ limit is shown by the dashed line. 
The difference between $eV_{\rm OC}$ and $E_{\rm g}$ corresponds to the difference between the optical band gap and 
the electronic band gap in the absence of dissociation energy. 
The difference originates from the distribution of carriers at the cell temperature. The experimental data for small molecules 
are denoted by closed circles and those for polymers are denoted by open circles. \cite{Lunt,Veldman}
The figure indicates that the smallest dissociation energy could be $0.3-0.4$ eV. 

\begin{figure}
\centerline{
\includegraphics[width=0.6\columnwidth]{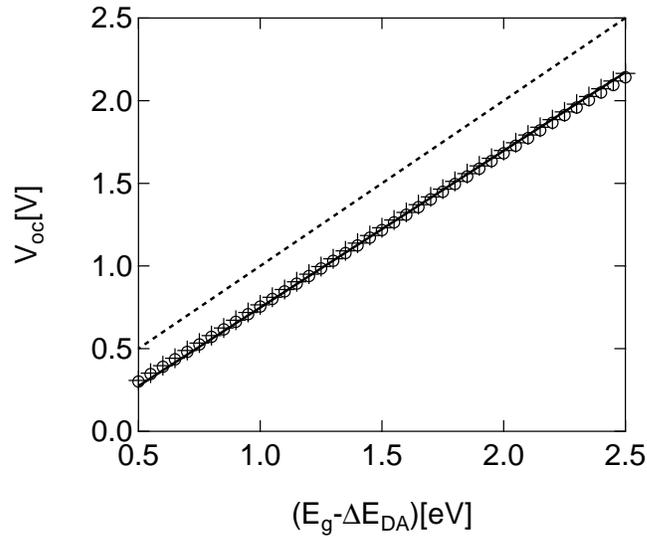}
}
\caption{$V_{\rm OC}$ calculated from Eq. (\ref{eq:3}) as a function of $E_{\rm g}-\Delta E_{\rm DA}$ (eV). 
The results for $\Delta E_{\rm DA}=0$ eV (SQ limit) and $\Delta E_{\rm DA}=0.4$ eV are shown by crosses and circles, respectively. 
The solid line indicates the relationship given by $eV_{\rm OC}   =E_{\rm g} -\Delta E_{\rm DA} -0.2 \mbox{(eV)}$. 
The dashed line indicates $eV_{\rm OC}=E_{\rm g}-\Delta E_{\rm DA}$. 
(The figure has been modified, with permission, from Ref. \citenum{surface}.)
}
\label{fig:3_1}
\end{figure}

When $e V_{\rm OC}$ (eV) values calculated from Eq. (\ref{eq:3}) are plotted 
against $E_{\rm g}-\Delta E_{\rm DA}$ (eV) 
for $\Delta E_{\rm DA}=0$, $0.3$ (not shown), and $0.4$ eV, 
they are close to the same linear line given by \cite{surface}
\begin{align}
\Delta E_{\rm DA} =E_{\rm g} -eV_{\rm OC}  -0.2 \mbox{(eV)} 
\label{eq:DeltaE}
\end{align}
as shown in Fig. \ref{fig:3_1}. 
$\Delta E_{\rm DA}$ can be estimated from $e V_{\rm OC}$ by this relation. 
It should be remembered that the relationship given by Eq. (\ref{eq:DeltaE}),  could be affected by changes in donor/acceptor ratio, 
layer thickness, and morphology. 


\section{Short-circuit current density}

\begin{figure}
\centerline{
\includegraphics[width=0.6\columnwidth]{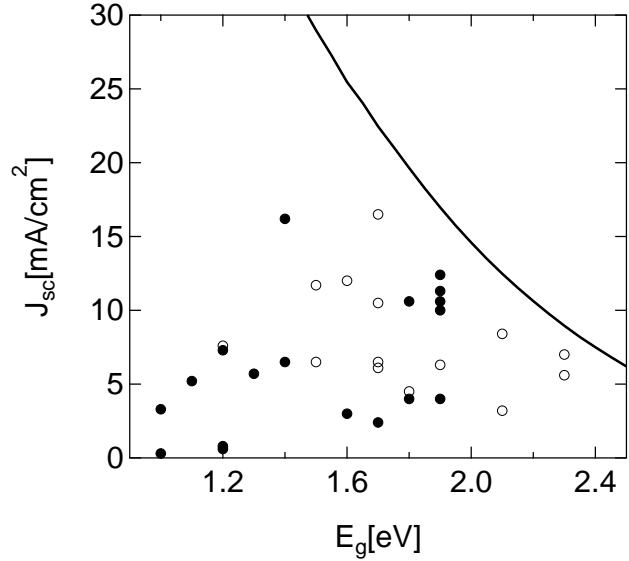}
}
\caption{Short circuit current density as a function of the optical band gap (eV) denoted by $E_{\rm g}$. 
The line indicates the theoretical result calculated using Eq. (\ref{eq:4}) with 
$\Delta E_{\rm DA}=0.4$ eV. 
The open and closed circles represent the experimental results obtained using small molecules and polymers, respectively. 
The experimental data are taken from Ref. \citenum{Lunt}.
}
\label{fig:4}
\end{figure}

As shown in Fig. \ref{fig:3}, 
the voltage loss to dissociate excitons into free carriers can be $0.3-0.4$ eV for some combinations of donors and acceptors. 
The corresponding maximum power conversion efficiency shown in Fig. \ref{fig:2}
is much higher than the current maximum power conversion efficiency,  
which is close to 11\%. 
To quantify the additional loss, 
we show the short-circuit current density calculated using Eq. (\ref{eq:1}) 
and compare it with the experimental data in Fig. \ref{fig:4}. 
The short-circuit current density was obtained from  
\begin{align}
J_{\rm sc} (E_{\rm g}, \Delta E_{\rm DA})=J_{\rm in}  
(E_{\rm g}  )-J_{\rm R}  (E_{\rm g}- \Delta E_{\rm DA} ,0). 
\label{eq:4}
\end{align}

Experimental values are much lower than the theoretical curve. 
The difference indicates the presence of current loss other than that caused by the radiative recombination characterized by 
black-body radiation. 
The short-circuit current is mainly given by $J_{\rm in}  (E_{\rm g}  )$, and the contribution of $J_{\rm R}  (E_{\rm g}- \Delta E_{\rm DA} ,0)$ is 
negligible 
for the the optical band gap (eV) shown in Fig. \ref{fig:4}. 
$J_{\rm R}  (E_{\rm g}- \Delta E_{\rm DA} ,0)$ is significant when $E_{\rm g}$ is close to $\Delta E_{\rm DA}$. 
It should also be considered that $J_{\rm R}  (E_{\rm g}- \Delta E_{\rm DA} ,V)$ may not be negligible 
under certain applied voltages. 
A large current loss is seen in Fig. \ref{fig:4} when  $\Delta E_{\rm DA} \leq 1.8$ eV, which is probably 
caused by nonradiative recombination. 

\section{Hybrid halide perovskite}
\begin{figure}
\centerline{
\includegraphics[width=0.3\columnwidth]{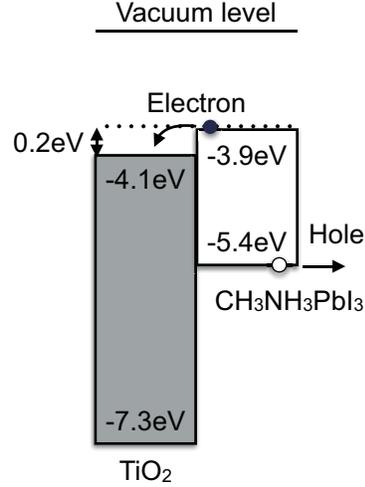}
}
\caption{Energy diagram of hybrid halide perovskite solar cells (CH$_3$NH$_3$PbI$_3$). }
\label{fig:5}
\end{figure}

Hybrid halide perovskite has been initially developed as a sensitizer of solar cells using electrolytes. \cite{Kojima,Im}
Recently, the power conversion efficiency of hybrid halide perovskite solar cells has rapidly been increased by avoiding electrolytes, and 
it has exceeded  
that of organic photovoltaics. \cite{Kim,Lee,Burschka,Liu,Heo}
A frequently studied material of hybrid perovskite solar cells is 
methylammonium lead iodide (CH$_3$NH$_3$PbI$_3$). 
The energy band diagram of CH$_3$NH$_3$PbI$_3$ is shown in Fig. \ref{fig:5}. \cite{Frost,Jung,Park}
The optical band gap of CH$_3$NH$_3$PbI$_3$ is $1.5$ eV. 
The dielectric constant of CH$_3$NH$_3$PbI$_3$ is around $25$, and the binding energy between an electron and a hole 
is small.  \cite{Frost}
The carriers dissociate in CH$_3$NH$_3$PbI$_3$, and 
the electron and hole lifetimes inside CH$_3$NH$_3$PbI$_3$ have been known to be large and are on the order of
$\mu$s. \cite{Ponseca,Stranks,Yamada,Shen,Yin}
The electron transfer from CH$_3$NH$_3$PbI$_3$ to TiO$_2$ occurs on a time scale of picoseconds, which 
is much faster than the time scale of charge recombination inside CH$_3$NH$_3$PbI$_3$. \cite{Shen,Han}
For simplicity, we assume that all electrons generated in CH$_3$NH$_3$PbI$_3$ transfer to TiO$_2$. 
This assumption may be reasonable for thin layers. 
The photon-absorbing layers of hybrid halide perovskite can be as thin as sub-$\mu$m 
because the optical absorption of CH$_3$NH$_3$PbI$_3$ is much higher than 
that of conventional inorganic semiconductors. \cite{Yin}

\begin{figure}
\centerline{
\includegraphics[width=0.6\columnwidth]{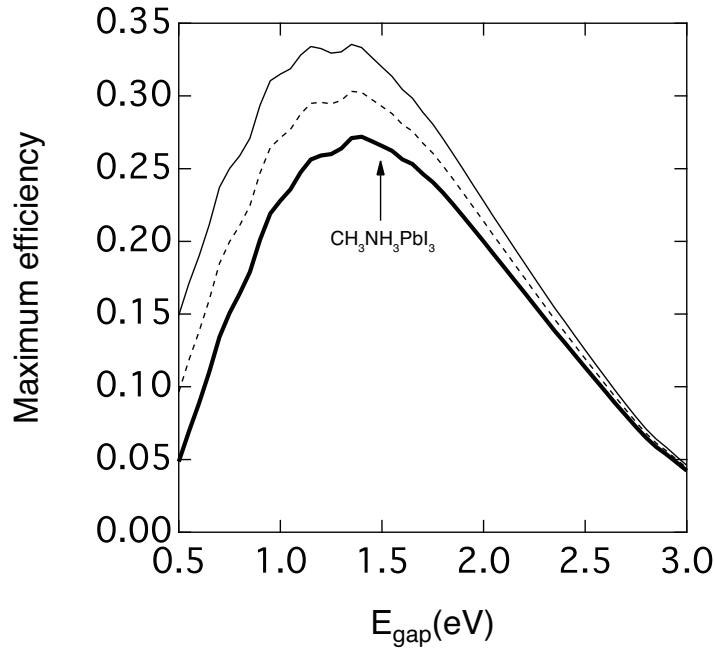}
}
\caption{Maximum power conversion efficiency as a function of the optical band gap (eV). The SQ limit is shown by the thin solid line. 
When electrons transfer from CH$_3$NH$_3$PbI$_3$ to TiO$_2$, 
the loss of energy is $0.2$ eV. The energy loss can be taken into account by the dissociation energy  
$\Delta E_{\rm DA}$ using Eq. (\ref{eq:2}). 
The thick solid and dashed lines are obtained using Eq. (\ref{eq:2}) with dissociation energies of 
$\Delta E_{\rm DA} =0.2$ and $0.1$ eV, respectively. 
The band gap and dissociation energy of CH$_3$NH$_3$PbI$_3$ are $1.5$ and $0.2$ eV, respectively.}
\label{fig:6}
\end{figure}

As seen in Fig. \ref{fig:5}, the transfer of electrons to TiO$_2$ is accompanied by an energy loss of $0.2$ eV.
According to recent reports, 
interface recombination between electrons in TiO$_2$ and holes in CH$_3$NH$_3$PbI$_3$ occurs on the nanosecond time scale. 
\cite{Shen}
In hybrid halide perovskites, 
Wannier excitons are formed and they dissociate inside hybrid halide perovskites. \cite{Frost}
When electrons transfer from hybrid halide perovskites to  TiO$_2$ with a time scale much shorter than 
the carrier lifetime and a thin layer of hybrid halide perovskites is used, 
charge recombination inside hybrid halide perovskites can be ignored. 
Charge recombination at the interface between hybrid halide perovskites and TiO$_2$ occurs 
on the nanosecond time scale and is taken into account.  
Although 
charge dissociation processes in hybrid halide perovskites are different from those in organic photovoltaic cells,  
which occur exclusively at the
donor-acceptor interfaces, 
the subsequent electron transfer and recombination processes in hybrid halide perovskites are virtually the same as 
those shown in Fig. \ref{fig:1}, where  
the energy loss associated with the transition of electrons from CH$_3$NH$_3$PbI$_3$ to TiO$_2$  is regarded as 
the dissociation energy denoted by $\Delta E_{\rm DA}$. 
The effect of energy loss on the maximum power conversion efficiency can be taken into account using 
Eq. (\ref{eq:2}) for the dissociation energy of 
$\Delta E_{\rm DA} =0.2$ [eV].
In Eq. (\ref{eq:2}), recombination of electrons in TiO$_2$ and holes in CH$_3$NH$_3$PbI$_3$ is assumed to be radiative. 
The results are shown by the thick solid line in Fig. \ref{fig:6}. 
 
The thick line at the optical band gap of $1.5$ eV corresponds to the results of CH$_3$NH$_3$PbI$_3$ with a TiO$_2$ layer. 
The maximum of the power conversion efficiency can be seen around the optical band gap of $1.4$ eV; the value is 
$26-27$ \%. 
Recently, a maximum efficiency of around $20$\% has been reported for a hybrid halide perovskite. \cite{Zhou,NREL,Jung}
According to Eq. (\ref{eq:DeltaE}), 
the open circuit voltage could be $1.1$ V when the optical band gap is $1.5$ eV and the energy loss is 
$0.2$ eV. 
The value is close to the experimental values. \cite{Jung}
The small energy loss of $0.2$ eV is consistent with 
the low operational losses for CH$_3$NH$_3$PbI$_3$, 
where the operational loss is defined as the difference between the absorption band gap and 
the maximum power voltage. \cite{Nayak}

The short-circuit current density of around $28$ mA/cm$^2$ 
at the optical band gap of $1.5$ eV can be obtained from Fig. \ref{fig:4}. 
The experimental values exceed $20$ mA/cm$^2$ and are smaller than $25$ mA/cm$^2$. \cite{Jung}
The difference indicates 
the existence of an additional current loss not taken into account in the theory. 

If the energy loss can be decreased to $0.1$ eV, 
the result is shown by the dashed line in Fig. \ref{fig:6}. 
The efficiency can be close to 30 \% when the optical band gap is $1.3-1.4$ eV.

At present, charge transfer and recombination processes in  hybrid halide perovskites are not clearly understood. 
If recombination takes place only inside a hybrid halide perovskite and the energy loss 
of $\Delta E$ occurs when electrons transfer from the hybrid halide perovskite to 
the electron transport layer, 
the proper expression should be changed from 
Eq. (\ref{eq:2}) to  
\begin{align}
Q(E_{\rm g}, \Delta E)=\frac{\mbox{Max} \left\{ (eV- \Delta E)\left[ J_{\rm in}  
(E_{\rm g}  )-J_{\rm R}  (E_{\rm g} ,V)\right]\right\}_V}{J_{\rm in} (0) } . 
\label{eq:5}
\end{align}
The result for $\Delta E=0.2$ eV, however, overlaps with the thick solid line in Fig. \ref{fig:6} (not shown).

\section{Conclusions}

The maximum power conversion efficiency is calculated by taking into account the dissociation energy of strongly bound charge pairs in organic materials. The dissociation energy can be estimated from the relationship between $eV_{\rm OC}$ and the optical band gap. 
The smallest possible values of $\Delta E_{\rm DA}$  are estimated as $0.3-0.4$ eV by analyzing the reported experimental data. When $\Delta E_{\rm DA} =0.4 $ eV, the peak value of the power conversion efficiency as a function of the optical band gap is theoretically given as 21\% at $827$ nm. The results support the advantage of using low band gap polymers for photon harvesting. In our calculations, nonradiative recombination of carriers is not considered. In direct band gap semiconductors used for inorganic PV cells, nonradiative recombination could be suppressed by reducing the number of defects. In organic PV cells, nonradiative recombination could be induced by electron-phonon coupling and is more difficult to suppress than in inorganic PV cells. Although we considered ideal organic PV, where nonradiative recombination is ignored, the results could be used  as a guide for selecting donor and acceptor materials.

Compared with OPV, recombination of carriers in hybrid halide perovskite has not yet been fully understood. 
By assuming that recombination takes place either at the interface between the hybrid halide perovskite  and TiO$_2$  
or inside  the hybrid halide perovskite, 
the theoretical limit of maximum power conversion efficiency was calculated in each case. 
The results were almost the same. 

When the energy loss associated with electron transfer from   the hybrid halide perovskite  to TiO$_2$  is 
$0.2$ eV, 
the optical band gap at the maximum power conversion efficiency is close to that of 
the hybrid halide perovskite. 
In this sense, the hybrid halide perovskite is an ideal light absorber. 
The difference between experimental values and the theoretical estimation is large for the 
short circuit current compared with that for the open-circuit voltage. 
The results suggest the possibility of increasing the power conversion efficiency by 
reducing the interface recombination. 

\acknowledgment
We would like to thank Dr. Kazaoui Said for discussions on the hybrid halide perovskite solar cells.

\end{document}